# Mimicking Surface Plasmons with Structured Surfaces

JB Pendry[1], L Martín-Moreno[2], and FJ Garcia-Vidal[3]

[1]Imperial College London, Dept. of Physics, The Blackett Laboratory,

London, SW7 2AZ, UK.

[2] Departamento de Fisica de la Materia Condensada, ICMA-CSIC, Universidad de

Zaragoza, E-50009 Zaragoza, SPAIN.

[3] Departamento de Fisica Teorica de la Materia Condensada, Universidad Autonoma de

Madrid, E-28049 Madrid, SPAIN.

Abstract

Metals such as silver support surface plasmons: electromagnetic surface excitations localised near the surface which originate from the free electrons of the metal. Surface modes are also observed on highly conducting surfaces perforated by holes. We establish a close connection between the two, showing that electromagnetic waves in both materials are governed by an effective permittivity of the same plasma form. Because the size and spacing of holes can readily be controlled on all relevant length scales, this gives the opportunity to create designer surface plasmons with almost arbitrary dispersion in frequency and in space, opening new vistas in surface plasmon optics.

The interaction of light with a metal surface is dominated by the free electrons that behave like a plasma with a dielectric function, $\varepsilon = 1 - \omega_p^2 / \omega^2$, which is negative below the plasma frequency, $\omega_p$. As a consequence metals support collective oscillations of the electrons bound to the surface (1). These are the surface plasmons, responsible for a host of phenomena unique to metals. Examples are: the surface enhanced Raman scattering (SERS) experiments (2) where the Raman signal is enhanced by the surface plasmon resonances sometimes as much as a factor of $10^6$, and the strong absorption of light in silver colloids (the black parts of photographic negatives). Interesting work on waveguiding in surface plasmon polariton band gap structures appears in (3) and plasmon propagation in metal stripes has been analysed in (4), whilst a study of 2D optics with surface plasmon polaritons can be found in (5). A useful summary of work appears in a recent review (6). More recently experiments have been performed on the transmission of light through sub-wavelength holes in metal films (7). It has been established (8) that resonant excitation of surface plasmons creates huge electric fields at the surface which force light through the holes, giving very high transmission coefficients.

Other materials with quite different dielectric functions from the plasma form can also support electromagnetic surface modes. A perfect conductor is an exception, but even such a material can be induced to support surface modes by drilling an array of holes in the surface. We show that these structured surfaces have many more properties in common with the electron plasma and that they can be described by an effective dielectric function of the plasma form, provided that the structure is on a scale much smaller than the wavelength of probing radiation. Reducing our description to a dielectric function of the plasma form has a conceptual elegance which unites all these phenomena under the same umbrella. An effective response generated by the structure of the medium is commonly seen at microwave frequencies (9-11). Thus it appears that surface structure may spoof surface plasmons which provide an accurate paradigm for structured surfaces. Such a description places the power of analogy in our hands unifying a broad range of phenomena. To illustrate these ideas we begin with a simple model. Let us suppose that we have a surface that is a perfect conductor pierced by an array of holes. For simplicity we shall assume a square cross section $a \times a$ for the holes and a square array of side $d$ (Fig. 1). We suppose that the holes and their spacing are much smaller that the wavelength of radiation, $a < d << \lambda_0$. An incident wave excites several waveguide modes in the holes but the fundamental mode will dominate because it is the least strongly decaying. Both electric and magnetic fields are zero inside the conductor but in the holes the electric field has the form,

$$\mathbf{E} = E_0 \begin{bmatrix} 0, & 1, & 0 \end{bmatrix} \sin(\pi x/a) \exp(ik_z z - i\omega t), \quad 0 < x < a, 0 < y < a \quad (1)$$

where

$$k_z = i\sqrt{\pi^2/a^2 - \varepsilon_h \mu_h k_0^2} \quad (2)$$

and $\varepsilon_h, \mu_h$ are the permittivity and permeability of any material that may be filling the holes. We take the *z*-axis to be the surface normal. The magnetic fields along the *x*- and *z*-axes follow from this equation.

Externally incident radiation is insensitive to the details of the holes, which it cannot resolve, and sees only an average response that we shall describe by $\varepsilon_z, \varepsilon_x = \varepsilon_y$ and $\mu_z, \mu_x = \mu_y$ for an anisotropic effective homogeneous medium. We can easily determine $\varepsilon_z, \mu_z$ by observing that the dispersion of the waveguide mode is unaffected by $k_x, k_y$ in either of the two possible polarisations and therefore,

$$\varepsilon_z = \mu_z = \infty \quad (3)$$

Suppose that the effective homogeneous fields in the medium are,

$$\mathbf{E}' = E'_0 [0, \ 1, \ 0] \exp(ik_x x + ik_z z - i\omega t) \quad (4)$$

We require that $k_z$ is the same as in the waveguide, and $k_x$ is defined by the incident direction. If this effective field is to match to the incident and reflected fields external to the surface, (1) and (4) must give the same average fields at the surface, hence after matching to incident and reflected waves,

$$\bar{E}_y = E_0 \frac{a}{d^2} \int_0^a \sin(\pi x/a) dx = E_0 \frac{2a^2}{\pi d^2} = E'_0 \quad (5)$$

We argue that the instantaneous flow of energy across the surface, $(\mathbf{E} \times \mathbf{H})_z$, must be the same inside and outside the surface, both for the real and effective media,

$$(\mathbf{E} \times \mathbf{H})_z = \frac{-k_z E_0^2}{\omega \mu_h \mu_0} \frac{a}{d^2} \int_0^a \sin^2(\pi x/a) dx = \frac{-k_z E_0^2}{\omega \mu_h \mu_0} \frac{a^2}{2d^2} =$$
$$(\mathbf{E}' \times \mathbf{H}')_z = \frac{-k_z E'^2_0}{\omega \mu_0 \mu_x} \quad (6)$$

Finer details of the fields around the holes are matched by strongly evanescent fields in the vacuum. Since these fields do not escape from the surface and contain little energy we neglect them. Substituting for $E'^2_0$ we deduce,

$$\mu_x = \mu_y = \frac{2d^2\mu_h}{a^2}\left[\frac{2a^2}{\pi d^2}\right]^2 = \frac{8a^2\mu_h}{\pi^2 d^2} \quad (7)$$

We also know that,

$$k_z = k_0\sqrt{\varepsilon_y \mu_x} = i\sqrt{\pi^2/a^2 - \varepsilon_h \mu_h k_0^2} \quad (8)$$

and hence,

$$\varepsilon_y = \varepsilon_x = \frac{1}{\mu_x}\left(\varepsilon_h\mu_h - \frac{\pi^2}{a^2 k_0^2}\right) = \frac{\pi^2 d^2 \varepsilon_h}{8a^2}\left(1 - \frac{\pi^2 c_0^2}{a^2 \omega^2 \varepsilon_h \mu_h}\right) \quad (9)$$

which is the canonical plasmon form with a plasma frequency of,

$$\omega_{pl} = \frac{\pi c_0}{a\sqrt{\varepsilon_h \mu_h}} \quad (10)$$

exactly the cut-off frequency of the wave guide.

It is easy to see that this is a general result: any array of closely spaced holes in a perfect conductor will have this form of effective response, with the plasma frequency given by the cut off frequency of the waveguide mode.

This effective medium implies a bound surface state when there is a divergence in the reflection coefficient of the surface for large values of $k_\parallel$ where $k'_z = i\sqrt{k_\parallel^2 - k_0^2}$ is imaginary. For the polarisation where the magnetic field is parallel to the surface,

$$r = \frac{k'_z - \varepsilon_\parallel^{-1} k_z}{k'_z + \varepsilon_\parallel^{-1} k_z} = \infty \quad (11)$$

which implies a dispersion relationship typical of a surface plasmon polariton,

$$k_\parallel^2 c_0^2 = \omega^2 + \frac{1}{\omega_{pl}^2 - \omega^2} \frac{64 a^4 \omega^4}{\pi^4 d^4} \qquad (12)$$

Figure 2 shows a sketch of the dispersion: note that at low frequencies the surface mode approaches the light line asymptotically, and the fields associated with the mode expand into the vacuum. At large $k_\parallel$ the frequency of the mode approaches $\omega_p$, in contrast to an isotropic plasma where the asymptote is $\omega_p/\sqrt{2}$. Although a flat perfectly conducting surface supports no bound states, the presence of holes, however small, produces a surface plasmon polariton-like bound surface state. In fact almost any disturbance of the flat surface will bind a state. Bound states are found in the following circumstances: a surface with an array of holes of finite depth however shallow; a surface with an array of grooves of finite depth; and a surface covered by a layer of dielectric, however thin.

It is also possible to produce hybrid surface plasmons by cutting holes in metals such as silver which already have a surface plasmon. The holes will increase the penetration of fields into the metal and lower the frequency of existing surface plasmons. In this case it is hard to distinguish between the 'real' and the spoof surface plasmons as they merge one into the other.

Our theory as presented is valid when the spacing between the holes is much less than the wavelength so that incident radiation cannot resolve the individual holes. However the theory can be progressively extended to larger hole spacings by including diffracted waves, more of which appear as the spacing gets bigger. We can do this by including Fourier components of ε and μ so that the effective medium also gives rise to diffracted beams. The main effect of diffraction is to couple the spoof plasmons to incident radiation.

Our result resolves a long standing debate over transmission through sub-wavelength holes. As mentioned above, when the anomalously high transmission first observed (7) it was attributed to resonant excitation of surface plasmons. Subsequent calculations (8) showed that sub-wavelength holes in a perfect conductor also gave rise to similar anomalous transmission even though the free surface of an unperforated conductor has no surface modes, leading to discussion as to the true origin of the effect. Now we can see that there are not two separate mechanisms: the holes will spoof surface plasmons which play the same resonant role as the real ones on silver.

The ability to engineer a surface plasmon at almost any frequency (metals are nearly perfect conductors from zero frequency up to the threshold of the THz regime) where none existed before, and to modify at will the frequency of surface plasmons on metals such as silver, opens opportunities to control and direct radiation at surfaces over a wide spectral range. Unlike other schemes for creating metamaterials with a plasma-like response (9-11) the present scheme is readily implemented, as hole drilling techniques are available over a range of length scales down to sub-optical wavelengths. Invoking a picture of the surface plasmon as a wave in two dimensions, we now have the ability to control the refractive index perceived by this wave. We stress that this is not the refractive index of the underlying material, but is given by the extent to which the surface plasmon wave vector deviates from the light line, $n_{sp} = k_{\|}/\omega_{sp}$. We have the option to design a lens to focus surface plasmons. Alternatively we could contour the refractive index, $n_{sp}(x,y)$, and adopting a particle picture of the surface plasmon, imagine it rolling like a ball over this landscape. More prosaic applications such as creating channels to act as waveguides are also possible.

Acknowledgements

We thank the EC under project FP6-NMP4-CT-2003-505699 for financial support. Additionally JB Pendry acknowledges support from the EPSRC, from DoD/ONR MURI grant N00014-01-1-0803, and from the EC Information Societies Technology (IST) programme Development and Analysis of Left-Handed Materials (DALHM), Project number: IST-2001-35511. L Martin-Moreno and FJ Garcia-Vidal also acknowledge financial support from the Spanish MCyT under contracts MAT2002-01534, MAT2002-00139 and BFM2003-08532-C02-01.

Mimicking Surface Plasmons with Structured Surfacespage 9

Figure captions

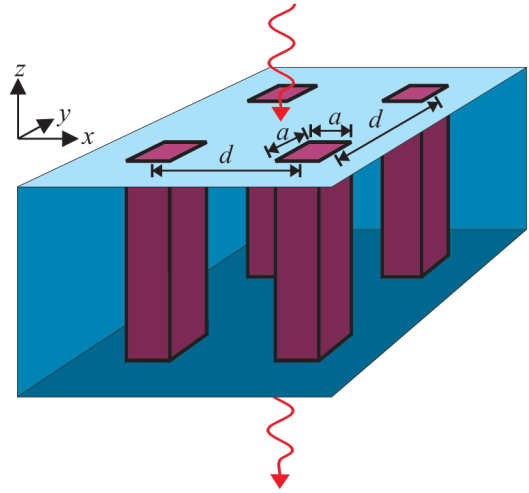

Figure 1
Our model system: $a \times a$ square holes arranged on a $d \times d$ lattice are cut into the surface of a perfect conductor. Our theory predicts localised surface plasmon modes induced by the structure.

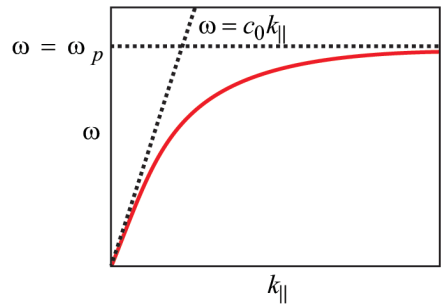

Figure 2.
Dispersion relation for spoof surface plasmons on a structured surface. Note the asymptotes of the light line at low frequencies, and the plasma frequency at large values of $k_{||}$.